\documentclass[reqno,11pt]{amsart}
\usepackage{amsmath,amssymb,tabularx,setspace,color,multirow,graphics,graphicx}
\newtheorem{theorem}{Theorem}
\usepackage[margin=3.5cm]{geometry}

\newtheorem{definition}{Definition}
\newtheorem{remark}{Remark}
\newtheorem{corollary}{Corollary}
\makeatletter
\renewcommand\section{\@startsection {section}{1}{\z@}%
	{-3.5ex \@plus -1ex \@minus -.2ex}%
	{2.3ex \@plus.2ex}%
	{\normalfont\large\bfseries}}
\makeatother

\newtheorem{Corollary}{Corollary}

\newtheorem{Theorem}{Theorem}

\newtheorem{Example}{Example}

\begin{document}
	\doublespace
	\vspace{-0.3in}
	\title[]{Goodness of fit tests for Rayleigh distribution }
\author[]%
{   V\lowercase{aisakh} K.   M.\textsuperscript{\lowercase{a}}, T\lowercase{homas} X\lowercase{avier} \textsuperscript{\lowercase{b} }\lowercase{and}
	~~S\lowercase{reedevi} E. P.\textsuperscript{\lowercase{c}}
	\\
	\lowercase{\textsuperscript{a}}S\lowercase{t.} T\lowercase{homas} C\lowercase{ollege}, T\lowercase{hrissur}, I\lowercase{ndia},\\
	\lowercase{\textsuperscript{b}}I\lowercase{ndian} S\lowercase{tatistical} I\lowercase{nstitute},
	D\lowercase{elhi}, I\lowercase{ndia},\\\lowercase{\textsuperscript{b}}M\lowercase{aharaja's} C\lowercase{ollege}, E\lowercase{rnakulam}, I\lowercase{ndia}.\\}
\thanks{{$^{\dag}$}{Corresponding E-mail: \tt sreedeviep@gmail.com}}
\maketitle
\vspace{-0.2in}

\begin{abstract}
	We develop a new goodness fit test for Rayleigh distribution for complete as well as right censored data. We use  U-Statistic theory  to derive the test statistic. First we develop a test for complete data and then discuss, how right censored observations can be incorporated in the testing procedure. The asymptotic properties of the test statistics in both uncensored and censored cases are studied in detail.  Extensive Monte Carlo simulation studies are carried out to validate the performance of the proposed tests. We illustrate the procedures using real data sets. We also provide, a goodness of fit test  for standard Rayleigh distribution based on jackknife empirical likelihood.
\end{abstract}


\section{Introduction}

Parametric approach for lifetime data often requires the assumption that the data follows a particular distribution. If this assumption about data can be proven to be right, it makes the further analysis easy. Goodness of fit tests are used to check whether the data follows a particular distribution. Exponential, gamma and Weibull distributions are some commonly used lifetime models. One other important lifetime distribution is Rayleigh distribution, which has its origin in an acoustics problem, where a study was conducted
on the resultant of a large number of sound waves with differing phases (Rayleigh, 1880). We can note that Rayleigh distribution is derived in literature as a special case of Weibull distribution with shape parameter 2. The distribution  has been implemented in the fields of astronomy, astrophysics and environmental sciences (see Fang and Margot (2012); Bovaird and Lineweaver (2017); Morgan et al. (2011); Celik (2004). The uses of Rayleigh distribution in survival analysis and reliability theory  is studied in Lawless (2011).

The goodness of fit tests for testing the hypothesis that the observed data follows Rayleigh distribution have been developed and studied. These include tests based on the empirical Laplace transform (Meintanis and Iliopoulos (2003)), entropy (Baratpour and Khodadadi (2013);  Noughabi et al. (2014)), the Hellinger distance (Jahanshahi et al. (2016)), a phi-divergence measure (Zamanzade and Mahdizadeh (2017)), moments (Best et al. (2010)) and the empirical likelihood ratio (Safavinejad et al. (2015)). The characterizations for Rayleigh distribution has  also been studied by many authors which include,  the conditional expectation characterization by Ahsanullah and Shakil (2013), the characterization based on record values by Nanda (2010) and the entropy based characterization by Baratpour and Khodadadi (2013). Recently, Ahrari et al. (2022) developed a quantile based test for testing Rayleigh distribution  and Liebenberg et al. (2022) proposed a new goodness of fit test for Rayleigh distribution based on conditional expectation characterization. However, all theses tests and studies were developed for complete data. Since, lifetime data involve censoring in most of the cases, it is important to develop a goodness of fit test for Rayleigh distribution, incorporating right censored observations. 

Stein (1972) introduced a natural identity for a random variable whose distribution belongs to an exponential family. Stein’s identity and its role in inference procedures have been discussed widely in the literature. For a detailed discussion on Stein’s type identity for a general class of probability distributions and related characterizations, one can refer to Sudheesh (2009) and Sudheesh and Dewan (2016) and the references therein. Ross (2011) discussed approximations of the Normal, Poisson, Exponential and Geometric distributions using Stein’s method.
As a special case,  let $X$ be a continuous random variable with finite mean $\mu$ and variance $\sigma^2$. Let $c(x)$ be a continuous function having first derivative. Then $X$ has normal distribution with mean $\mu$ and variance $\sigma^2$ if and only if
\begin{equation*}\label{stein}
	E(c(X)(X-\mu))=\sigma^2E(c'(X)),
\end{equation*}
This has come to be known in literature as Stein’s identity or Stein’s lemma.
Using Stein’s type identity, Betsch and Ebner (2019) developed a fixed point characterization for gamma distribution. Making use of this characterization Vaisakh et al.(2021), developed a U-statistic based goodness fit test for gamma distribution  for complete data as well as censored data. We use Stein’s type identity for Weibull distribution, to develop a goodness of fit for Rayleigh distribution. Then we discuss, how the testing procedure can be extended to incorporate right censored observations. 

The rest of the article is organized as follows. In Section 2, making use of the Stein's type identity for Weibull distribution, we develop a U-statistic based goodness fit test for Rayleigh distribution. In Section 3 we discuss how to incorporate right censored observations. Extensive Monte Carlo Simulation studies are carried out in Section 4 to assess the performance of the test in finite sample. The illustrations of the proposed methods using real data sets is given in and data analysis are reported in Section 5. A jackknife empirical likelihood ratio  test is developed for standard Rayleigh distribution in Section 6. Finally, we conclude the study in Section 7.

\section{Test statistic: Uncensored case}
In this section, we develop a goodness of fit test for Rayleigh  distribution.  We use fixed point characterization based on Stein's type identity for Rayleigh distribution to develop the test.
\begin{theorem}
	The random variable $X$ has Rayleigh distribution with parameter $\sigma$ if and only if 
	$$F(t)=E\left[\left(\frac{x^2-\sigma^2}{x\sigma^2}\right) \min(X,t)\right].$$
\end{theorem}
\vspace{-1cm}
Based on a random sample $X_{1}, ...,X_{n}$  from $F$, we are interested in  testing the null hypothesis
$$H_{0}: \text{$F$  has Rayleigh distribution}.$$
against$$ H_1: \text{$F$ does not follow Rayleigh distribution}.$$
For testing the above hypothesis first we define a departure measure which discriminate between null and alternative hypothesis. Consider  $ \Delta(F)$ given by
\begin{eqnarray}\label{deltam}
	\Delta(F)&=&\int_{0}^{\infty}\left(E\left[\left(\frac{x^2-\sigma^2}{x\sigma^2}\right) min(X,t)\right]-F(t)\right)dF(t).
\end{eqnarray}In view of Theorem 1, $\Delta(F)$ is zero under  $H_0$ and not zero under $H_1$. Hence $\Delta(F)$  can be considered as a measure of departure  from the null hypothesis $H_0$ towards the alternative  hypothesis $H_1$.

As we developed  the  test using the theory of U-statistics, first we simplify $\Delta(F)$ in terms of expectation of the function of random variables. Consider
\begin{eqnarray}\label{delta}
	\Delta(F)&=&\int_{0}^{\infty}E\left[\left(\frac{x^2-\sigma^2}{x\sigma^2}\right) min(X,t)\right]dF(t)-\int_{0}^{\infty}F(t)dF(t).\nonumber\\
	&=&\int_{0}^{\infty}\int_{0}^{\infty}\left(\frac{x^2-\sigma^2}{x\sigma^2}\right)min(x,t)dF(x)dF(t)- 1/2\nonumber\\
	&=&\frac{1}{\sigma^2}\int_{0}^{\infty}\int_{0}^{t}\left({x^2-\sigma^2}\right)dF(x)dF(t)\nonumber\\&&\qquad+\frac{1}{\sigma^2}\int_{0}^{\infty}\int_{t}^{\infty}\left({x^2-\sigma^2}\right)\frac{t}{x}dF(x)dF(t)- 1/2\nonumber\\&=&\frac{1}{\sigma^2}\int_{0}^{\infty}\int_{0}^{t}x^2dF(x)dF(t)\nonumber\\&&\qquad+\frac{1}{\sigma^2}\int_{0}^{\infty}\int_{t}^{\infty}\left({x^2-\sigma^2}\right)\frac{t}{x}dF(x)dF(t)- 1\\\nonumber
	&=&\frac{1}{\sigma^2}(\Delta_1(F)+\Delta_2(F))-1 \,\,(say).
\end{eqnarray}
Now, changing the order of integration we have
\begin{eqnarray}\label{delta1}
	\Delta_1(F)&=&\int_{0}^{\infty}\int_{0}^{t}x^2dF(x)dF(t)\nonumber\\
	&=&\int_{0}^{\infty}{x^2}\int_{x}^{\infty}dF(t)dF(x)\nonumber\\
	&=&\int_{0}^{\infty}{x^2}\bar F(x)dF(x)\nonumber\\
	&=&\frac{1}{2}E\left(\min(X_{1},X_{2})^2\right).
\end{eqnarray}Again
\begin{eqnarray}\label{delta2}
	\Delta_2(F)&=&\int_{0}^{\infty}\int_{t}^{\infty}\left({x^2-\sigma^2}\right)\frac{t}{x}dF(x)dF(t)\nonumber\\
	&=&\int_{0}^{\infty}\int_{t}^{\infty}{t}{x}dF(x)dF(t)-\sigma^2\int_{0}^{\infty}\int_{t}^{\infty}\frac{t}{x}dF(x)dF(t)\nonumber\\
	&=&E\left((X_1X_2)I(X_2<X_1)\right)-\sigma^2E\left(\frac{X_2}{X_1}I(X_2<X_1)\right)
\end{eqnarray}
Substituting (\ref{delta1}) and (\ref{delta2}) in (\ref{delta}), we obtain
\begin{eqnarray}\label{deltafinal}
	\Delta(F)=\frac{1}{\sigma^2}E\left(\min(X_{1},X_{2})^2+X_1X_2I(X_2<X_1)\right)-E\left(\frac{X_2}{X_1}I(X_2<X_1)\right)-1.
\end{eqnarray}

We find test statistic using  U-statistics theory. For Rayleigh distribution
$$E(X^2)=2\sigma^2.$$ Hence an unbiased and consistent  estimator of $\sigma^2$ is given by
$$\widehat\sigma^2=\frac{1}{2n}\sum_{i=1}^{n}X_i^2.$$
Defined a symmetric kernel
\begin{eqnarray*}
	h_1(X_1,X_2) &=&\frac{1}{2} \left(2\min(X_{1},X_{2})^2+{X_{1}}{X_{2}}I(X_{2}<X_{1})+{X_{1}}{X_{2}}I(X_{2}<X_{1})\right) \\&=&\frac{1}{2}(2\min(X_{1},X_{2})^2+X_1X_2).
\end{eqnarray*}
Then a U-statistic defined by
$$U_1=\frac{2}{n(n-1)}\sum_{i=1}^{n}\sum_{j=1,j<i}^{n}h_1(X_i,X_j),$$ is an unbiased estimator of $E\left(\min(X_{1},X_{2})^2+{X_{1}}{X_{1}}I(X_{2}<X_{1})\right)$.
Again, consider a symmetric kernel
$$h_2(X_1,X_2) =\frac{1}{2} \left(\frac{X_{1}}{X_{2}}I(X_{1}<X_{2})+\frac{X_{2}}{X_{1}}I(X_{2}<X_{1})\right) .$$ Then a U-statistic defined by
$$U_2=\frac{2}{n(n-1)}\sum_{i=1}^{n}\sum_{j=1,j<i}^{n}h_2(X_i,X_j),$$ is an unbiased estimator of $E\left(\frac{X_{2}}{X_{1}}I(X_{2}<X_{1})\right)$.
Hence the test statistic is given by
\begin{equation}
	\widehat{\Delta} =\frac{U_{1}}{\widehat{\sigma}^2}-U_{2}-1.
\end{equation}
Let $X_{(i)}$ be the order statistics based on $n$ random sample $X_1,\ldots,X_n$ from $F$. Then in terms of order statistics we can rewrite the test statistics as
$$\widehat{\Delta}=\frac{2}{n(n-1)\widehat\sigma^2}\sum_{i=1}^{n}(n-i)X_{(i)}^2+
\frac{1}{n(n-1)}\sum_{j=1}^{n-1}\sum_{i=j+1}^{n}X_{(j)}\left(\frac{X_{(i)}}{\widehat\sigma^2}-\frac{1}{X_{(i)}}\right)-1.$$
We reject the null hypothesis $H_0$ against the alternative  $H_1$ for large value of $\widehat{\Delta}$.

Next we study the asymptotic properties of the test statistic.  Since $\widehat\sigma^2$, $U_1$ and $U_2$ are U-statistics they are consistent estimators of $\sigma$, $E(\min(X_{1},X_{2}))$ and $E\left(\frac{X_{1}}{X_{2}}I(X_{1}<X_{2})\right)$, respectively (Lehmann, 1951). Hence  the following result is straight forward.
\begin{theorem}Under $H_1$, as $n\rightarrow \infty$,  $\widehat{\Delta}$ converges in probability to ${\Delta}(F)$.
\end{theorem}
\begin{theorem}
	As $n\rightarrow \infty$,  $\sqrt{n}(\widehat{\Delta}-\Delta(F))$ converges in distribution to normal random variable with mean zero and variance $\sigma^2$, where $\sigma^2$ is given by
	\begin{small}
		\begin{equation}\label{var}
			\sigma^{2}=Var\Big(\frac{2X^2\bar F(X)}{\sigma^2} +\frac{2}{\sigma^2}\int_{0}^{X}y^2dF(y)+\frac{X\mu}{\sigma^2}-X\int_{X}^{\infty}\frac{1}{y}
			dF(y)-\frac{1}{X}\int_{0}^{X}ydF(y)\Big).
		\end{equation}
	\end{small}
\end{theorem}
\noindent {\bf Proof:}
Define $$\widehat{\Delta}^{*}(F) =\frac{U_{1}}{\sigma^2}-U_{2}.$$  Since $\widehat \sigma^2$ is a consistent estimators of $\sigma^2$, by Slutsky's theorem,  the  asymptotic distribution of $\sqrt{n}(\widehat{\Delta}-\Delta(F))$  and  $\sqrt{n}(\widehat{\Delta}^{*}-E(\widehat{\Delta}^{*}))$ are same. Now we observe  that $\widehat{\Delta}^*$ is a U statistic with symmetric kernel,
\begin{small}
	$$h(X_1,X_2)  = \frac{1}{2}\left(\frac{2\min(X_{1},X_{2}^2)}{\sigma^2}+\frac{X_1X_2}{{\sigma^2}}-
	\frac{X_{1}}{X_{2}}I(X_{1}<X_{2})-\frac{X_{2}}{X_{1}}I(X_{2}<X_{1})\right).$$
\end{small}
Hence using the central limit theorem for U-statistics  we have the asymptotic  normality of $ \widehat{\Delta}^*$. The asymptotic variance is $4\sigma_1^2$ where $\sigma_1^2$ is given by (Lee, 2019)
\begin{equation}\label{var1}
	\sigma_1^2= Var\left[E\left(h(X_{1},X_{2})|X_{1}\right)\right].
\end{equation}Consider
\begin{eqnarray}\label{var11}
	E[2\min(x,X_{2})^2+xX_2]&=&2E[x^2 I(x<X_{2})+X_2^2I(X_{2}<x)+xX_2]\nonumber \\
	&=&2x^2P(x<X_2)+2\int_{0}^{\infty}y^2I(y<x)dF(y)+x\mu\nonumber\\
	&=&2x^2\bar F(x)+2\int_{0}^{x}y^2dF(y)+x\mu.
\end{eqnarray}Also
\begin{eqnarray}\label{var12}
	E[\frac{x}{X_2}I(x<X_2]+E[\frac{X_2}{x}I(X_2<x)] &=&xE[\frac{1}{X_2}I(x<X_2]+\frac{1}{x}E[X_2I(X_2<x)]\nonumber\\
	&=&x\int_{0}^{\infty}\frac{1}{y}I(x<y)dF(y)+\frac{1}{x}\int_{0}^{x}ydF(y)\nonumber\\
	&=&x\int_{x}^{\infty}\frac{1}{y}dF(y)+\frac{1}{x}\int_{0}^{x}ydF(y).
\end{eqnarray}
Substituting equations (\ref{var11}) and (\ref{var12})  in equation  (\ref{var1}) we obtain the variance expression as specified in the theorem.

Under the null hypothesis $H_0$, $\Delta{(F)}=0$. Hence we have the following corollary.
\begin{corollary}
	Under $H_0$, as $n\rightarrow \infty$,  $\sqrt{n}\widehat{\Delta}$ converges in distribution to normal with mean zero and variance $\sigma_0^2$, where $\sigma_0^2$ is the value of $\sigma^2$ evaluated under $H_0$ and it is given by
	$$\sigma_0^2=\frac{1}{4}E\left(8 - 6 e^{\frac{-X^2}{2s^2}}- \frac{\sqrt{2\pi}(s^2-X^2)erf\left[\frac{X}{\sqrt{2}s}\right]}{sX}\right)^2-9,$$
	where $erf(x)=\frac{2}{\sqrt{\pi}}\int_{0}^{x}e^{-t^2}dt$.
	
	\end{corollary}
	
	An asymptotic critical region of the test can be obtain  using Corollary 1. Let $\widehat\sigma_{0}^2$ be a consistent estimator of $\sigma_{0}^2$. We reject the null hypothesis $H_{0}$ against the alternative hypothesis $H_{1}$ at a significance level $\alpha$, if
	\begin{equation*}
		\frac{ \sqrt{n} |\widehat{\Delta}| }{\widehat\sigma_0}>Z_{\alpha/2},
	\end{equation*}
	where $Z_{\alpha}$ is the upper $\alpha$-percentile point of the standard normal distribution.
	
	Finding a  consistent estimator of the null variance $\sigma_{0}^2$ is difficult. Hence we find the  critical region of the proposed test using Monte Carlo simulation. We determine lower ($c_1$) and upper ($c_2$)  quantiles in such a way that $P(\widehat\Delta<c_1)=P(\widehat\Delta>c_2)=\alpha/2$.  Finite sample performance of the  test is evaluated through Monte Carlo simulation study and the results are presented in Section 4.

	\section{Test statistic: Censored case}
	Next we discuss how the right censored observations can be incorporated in  the proposed testing  method.
	Consider the right-censored  data  $(Y, \delta)$, with $Y=\min(X,C)$ and $\delta=I(X\leq C)$, where $C$ is the censoring time. We assume censoring times and lifetimes are independent. Now we are interested to test the hypothesis discussed  in Section 2 based on $n$ independent and identical observation $\{(Y_{i},\delta_i),1\leq i\leq n\}$.  As we developed the test based on  U-statistics for right censored data, we use the same  departure measure $\Delta (F)$ given in (\ref{deltam}). For that purpose, we rewrite (\ref{deltafinal}) as
	\begin{small}
		\begin{eqnarray*}\label{deltamcens}
			\Delta(F)=E\left(\frac{\min(X_{1},X_{2})^2}{\sigma^2}+\frac{X_1X_2I(X_2<X_1)}{\sigma^2}-\frac{X_{2}}{X_{1}}I(X_{2}<X_{1})-2I(X_2<X_1)\right).
		\end{eqnarray*}
	\end{small}
	To develop the test statistic for right censored case, we estimate each quantity in $\Delta(F)$ using U-statistics for right censored data (Datta et al., 2010). An estimator of $E(\min(X_{1},X_{2})^2+X_1X_2I(X_2<X_1))$ is given by
	\begin{equation}\label{delta1c}
		\widehat{\Delta}_{1c}=\frac{1}{n(n-1)}\sum_{i=1}^{n}\sum_{j<i;j=1}^{n}\frac{(2\min(Y_{1},Y_{2})^2+X_1X_2)\delta_i\delta_j}{\widehat{K}_{c}(Y_i)\widehat{K}_{c}(Y_j)},
	\end{equation}
	provided $\widehat{K}_{c}(Y_i)>0$ and $\widehat{K}_{c}(Y_j)>0$, with probability 1 and $\widehat{K}_c$ is the Kaplan-Meier estimator of $K_c$, the survival function of  $C$.
	Again, an estimator of   $E\left(\frac{X_{2}}{X_{1}}I(X_{2}<X_{1})+2I(X_2<X_1)\right)$ is given by
	\begin{equation}\label{delta2c}
		\widehat{\Delta}_{2c}=\frac{1}{n(n-1)}\sum_{i=1}^{n}\sum_{j<i;j=1}^{n}\frac{(\frac{X_{i}}{X_{j}}I(X_{i}<X_{j})+\frac{X_{j}}{X_{i}}I(X_{j}<X_{i})+2)\delta_i\delta_j}{\widehat{K}_{c}(Y_i)\widehat{K}_{c}(Y_j)}.
	\end{equation}
	Similarly, an estimators of $\sigma^2=E(X^2)/2$ is given by
	\begin{equation}\label{xbarc}
		\sigma_{c}^2=\frac{1}{2n}\sum_{i=1}^{n}\frac{Y_i^2 \delta_i}{\widehat{K}_{c}(Y_i)}.
	\end{equation}
	Using the estimators given in equations (\ref{delta1c}-\ref{xbarc}), we obtain the test statistic as
	\begin{equation}\label{esticen}
		\widehat{\Delta}_{c}=\frac{\widehat{\Delta}_{1c}}{\widehat{\sigma}_{c}^2}-\widehat{\Delta}_{2c}.
	\end{equation} Hence in the right censored case, we reject $H_{0}$ in favour of $H_{1}$ for large values of $\widehat{\Delta}_c$.
	
	To obtain the limiting distribution of $ \widehat{\Delta}_{c}$, let $N_i^c(t)=I(Y_i\leq t, \delta_i=0)$ be the counting process corresponds to the censoring variable $C_i$. Denote $R_i(t)=I(Y_i\geq t)$. Also let $\sigma^2_c$ be the  hazard rate of $C$.  The martingale associated with this counting process $N_i^c(t)$ is given by
	\begin{equation*}
		M_i^c(t)=N_i^c(t)-\int_{0}^{t} R_i(u) \sigma^2_c(u) du.
	\end{equation*}
	Let $G(x,y)=P(X_{1}\leq x, Y_{1}\leq y,\,  \delta=1),\, x\in \mathcal{X}$, $\bar H(t)=P(Y_{1}> t)$ and
	\begin{equation*}
		w(t)=\frac{1}{\bar{H}(t)} \int_{\mathcal{X}\times[0,\infty)}{\frac{h_1(x)}{K_c(y-)}I(y>t)dG(x,y)},
	\end{equation*}
	where $h_1(x)=E(h(X_1,X_2)|X_1=x).$  The proof of next result follows from Theorem 1 of Datta et al. (2010) for a particular choice of the kernel.
	\begin{theorem}\label{thm5.4}
		Let \begin{small}
			$$h_1(x)  =\frac{1}{2}E\left(\frac{2\min(x,Y_{2})^2+xX_2}{\sigma^2}-\frac{x}{Y_{2}}I(x<Y_{2})-\frac{Y_{2}}{x}I(Y_{2}<x)-2\right)$$\end{small}
		Suppose the conditions \begin{small} $$E\left[\left(\frac{2\min(Y_{1},Y_{2})^2+X_1X_2}{\sigma^2}\right)-
			E\left(\frac{Y_{1}}{Y_{2}}I(Y_{1}<Y_{2})+\frac{Y_{2}}{Y_{1}}I(Y_{2}<Y_{1})+2\right)\right]^2 <\infty,$$\end{small} $\int_{\mathcal{X}\times[0,\infty)}{\frac{h_1^{2}(x)}{K_c^2(y)}dG(x,y)}<\infty$ and  $\int_0^\infty w^2(t)\sigma^2_c(t)dt<\infty$ holds.\\ As $n \rightarrow \infty $,  $\sqrt{n}(\widehat{\Delta}_c-\Delta(F))$ converges in  distribution to a Gaussian random variable with mean zero and variance $4\sigma_{c}^{2}$, where $\sigma_{c}^2$  is given by
		\begin{equation*}
			\sigma_{c}^{2}=Var\Big(\frac{h_1(X)\delta_1}{K_c(Y_1-)}+\int w(t) dM_1^c(t)\Big).
		\end{equation*}
	\end{theorem}
	
	Next we find an estimator of  $\sigma_{c}^{2}$ using the reweighed techniques. An estimator of $\sigma_{c}^2$ is given by
	\begin{equation*}\label{ecvar}
		\widehat{\sigma}_{c}^2=\frac{4}{(n-1)}\sum_{i=1}^{n}(V_{i}-\bar V)^2,
	\end{equation*}
	where
	\begin{equation*}\label{36}
		V_{i}=\frac{\widehat{h}_1(X_{i})\delta_i}{\widehat{K}_c(Y_{i})}+\widehat w(X_{i})(1-\delta_i)-\sum_{j=1}^{n}\frac{\widehat w(X_{i})I(X_{i}>X_{j})(1-\delta_i)}{\sum_{i=1}^{n}I(X_{i}>X_{j})},
	\end{equation*}
	$$\bar V =\frac{1}{n}\sum_{i=1}^{n}V_{i}, \quad  \widehat{h}_1(X)=\frac{1}{n}\sum_{i=1}^{n}\frac{h(X,Y_{i})\delta_i}{\widehat{K}_c(Y_{i}-)}, \quad R(t)=\frac{1}{n}\sum_{i=1}^{n}I(Y_{i}>t)$$ and
	$$\widehat w (t)=\frac{1}{R(t)}\sum\limits_{i=1}^{n}\frac{\widehat{h}_1(X_{i})
		\delta_{i}}{\widehat{K}_c(Y_{i})}I(X_{i}>t).$$

	Let $\widehat{\sigma}_{0c}^2$ be the value of $\widehat{\sigma}_{c}^2$ evaluated under $H_0$. Under right censored situation, we reject the null hypothesis $H_{0}$ against the alternative hypothesis $H_{1}$ at a significance level $\alpha$, if
	\begin{equation*}
		\frac{ \sqrt{n} |\widehat{\Delta}_c| }{\widehat{\sigma}_{0c}}>Z_{\alpha/2}.
	\end{equation*}
	The results of the Monte Carlo simulation which assess the finite sample performance of the test  is also reported in Section 4.
	%
	%
	\vspace{-1cm}
	\section{Empirical evidence}
	The finite sample performance of the proposed test procedure is evaluated through  a Monte Carlo simulation study using R software. To show the competitiveness of our test with the existing test procedures for complete data, we compare the empirical powers of the same. In censored case we evaluate the power of our test against different alternatives at different censoring percentage.
	
	\subsection{Uncensored case}
	
	First we find the empirical type I error and empirical power of the  proposed test and other  tests used for comparison.   The algorithm used for finding the empirical power can be summarised as follows;
	\begin{enumerate}
		\item Generate lifetime data from the desired alternative and calculate the test statistic.
		\item Find an estimator $\widehat\sigma^2$ of the parameter $\sigma^2$ using the data obtained in Step 1.
		\item Generate a sample of size $n$ from Rayleigh distribution with parameter $\widehat\sigma^2$.
		\item Repeat Step 3, 10000 times and determine the simulated critical points. 
		\item Repeat Steps 1-4 10000 times and calculate the empirical power as the proportion of significant test statistics.
	\end{enumerate}
	First we find empirical type I error of the test. We generate observation from standard Rayleigh distribution, with different samples sizes $n = $ 10, 20, 30, 40 and 50 to calculate the empirical type I error. To find the empirical  power, lifetime random variables are generated from different choices of alternative  including Weibull, Gamma, lognormal, Pareto and half-normal distributions. 
	
	We compare the performance of our test with goodness of fit test for Rayleigh distribution proposed by Ahrari et al. (AH) (2022), Meintanis and Illipoulos (MI) (2003), Baratpour and Khodadadi (BK) (2013), Jahanshahi et al.(JH) (2016) and also with the well-known  Kolmogorov Smirnov (KS) test and Cramer von Mises (CvM) test. 
	We report, the results of the simulation study for empirical type I error in Table \ref{Tab:1}. We can see that, in Table \ref{Tab:1} the size of the test attains chosen level of significance. Table $\ref{Tab:2}$ suggests that the developed test has good power against all choices of alternatives which increases with sample size. We can see that the newly proposed test performs better than other tests in almost all of the cases we considered. The proposed test has  high power even for small sample size, which affirms the efficiency of the test. When the alternative has a Pareto distribution, then KS test and CvM test perform better than the developed test.
	
	\begin{table} 
		\caption{Comparisons of empirical type I error}
		\label{Tab:1} 
		\begin{tabular}{lllllllll} \hline \noalign{\smallskip}
			$n$ & $\alpha$ & $\hat{\Delta}$ & KS & CvM & AH & MI & BK & JH\\
			\noalign{\smallskip} \hline \noalign{\smallskip}
			10 & 0.01 & 0.0104 & 0.0183 & 0.0188 & 0.0093 & 0.0133 & 0.0101 & 0.0096 \\
			20 &  & 0.0107 & 0.0136 & 0.0150 & 0.0105 & 0.0076 & 0.0094 & 0.0101 \\
			30 &  & 0.0113 & 0.0125 & 0.0137 & 0.0105 & 0.0109 & 0.0086 & 0.0083 \\
			40 &  & 0.0092 & 0.0110 & 0.0110 & 0.0092 & 0.0093 & 0.0104 & 0.0096 \\
			50 &  & 0.0090 & 0.0107 & 0.0112 & 0.0096 & 0.0077 & 0.0107 & 0.0106 \\
			\noalign{\smallskip} \hline \noalign{\smallskip}
			10 & 0.05 & 0.0507 & 0.0638 & 0.0627 & 0.0478 & 0.0495 & 0.0489 & 0.0500 \\
			20 &  & 0.0477 & 0.0582 & 0.0597 & 0.0446 & 0.0486 & 0.0516 & 0.0503 \\
			30 &  & 0.0532 & 0.0526 & 0.0557 & 0.0480 & 0.0536 & 0.0550 & 0.0538 \\
			40 &  & 0.0494 & 0.0550 & 0.0545 & 0.0502 & 0.0492 & 0.0476 & 0.0483 \\
			50 &  & 0.0472 & 0.0493 & 0.0487 & 0.0487 & 0.0510 & 0.0497 & 0.0542 \\
			\noalign{\smallskip} \hline \noalign{\smallskip}
		\end{tabular} 
	\end{table}
	
	\begin{table} 
		\footnotesize
		\caption{Comparisons of empirical power}
		\label{Tab:2} 
		\begin{tabular}{llllllllll} \hline \noalign{\smallskip}
			& $n$ & $\alpha$ & $\hat{\Delta}$ & KS & CvM & AH & MI & BK & JH\\
			\noalign{\smallskip} \hline \noalign{\smallskip}
			Weibull (1.5) & 10 & 0.01 & 0.2257 & 0.0080 & 0.0015 & 0.0518 & 0.0155 & 0.0480 & 0.0767 \\
			& 20 &  & 0.6792 & 0.0409 & 0.0211 & 0.2666 & 0.0533 & 0.0976 & 0.5245 \\
			& 30 &  & 0.8952 & 0.1048 & 0.0882 & 0.5584 & 0.1280 & 0.1451 & 0.7358 \\
			& 40 &  & 0.9782 & 0.2216 & 0.2403 & 0.8207 & 0.1738 & 0.2768 & 0.8840 \\
			& 50 &  & 0.9953 & 0.3626 & 0.4622 & 0.9298 & 0.9429 & 0.4220 & 0.9119 \\
			\noalign{\smallskip} \hline \noalign{\smallskip}
			& 10 & 0.05 & 0.5089 & 0.0704 & 0.0481 & 0.2365 & 0.0493 & 0.1463 & 0.3287 \\
			& 20 &  & 0.8931 & 0.2084 & 0.2232 & 0.6550 & 0.1294 & 0.2018 & 0.8849 \\
			& 30 & & 0.9743 & 0.4181 & 0.5205 & 0.8812 & 0.3971 & 0.3282 & 0.9512 \\
			& 40 & & 0.9975 & 0.6145 & 0.7610 & 0.9666 & 0.9595 & 0.5759 & 0.9745 \\
			& 50 & & 0.9992 & 0.8027 & 0.9093 & 0.9923 & 0.9694 & 0.7971 & 0.9878 \\
			\noalign{\smallskip} \hline \noalign{\smallskip}
			Gamma (1,1) & 10 & 0.01 & 0.2794 & 0.0066 & 0.0022 & 0.0545 & 0.0644 & 0.0354 & 0.1965 \\
			& 20 &  & 0.6169 & 0.0409 & 0.0225 & 0.2782 & 0.1018 & 0.2207 & 0.7462 \\
			& 30 &  & 0.8893 & 0.1116 & 0.0940 & 0.6653 & 0.2948 & 0.6869 & 0.9250 \\
			& 40 &  & 0.9831 & 0.2222 & 0.2475 & 0.8529 & 0.4564 & 0.9030 & 0.9799 \\
			& 50 &  & 0.9949 & 0.3598 & 0.4623 & 0.9462 & 0.7129 & 0.9818 & 0.9913 \\
			\noalign{\smallskip} \hline \noalign{\smallskip}
			& 10 & 0.05 & 0.5192 & 0.4583 & 0.4714 & 0.2388 & 0.3475 & 0.1823 & 0.3204 \\
			& 20 &  & 0.8741 & 0.8160 & 0.8512 & 0.6599 & 0.6432 & 0.6523 & 0.8739 \\
			& 30 &  & 0.9787 & 0.9540 & 0.9701 & 0.8886 & 0.8296 & 0.9307 & 0.9480 \\
			& 40 &  & 0.9972 & 0.9910 & 0.9957 & 0.9658 & 0.9421 & 0.9853 & 0.9775 \\
			& 50 &  & 0.9993 & 0.9988 & 0.9997 & 0.9949 & 0.9761 & 0.9958 & 0.9859 \\
			\noalign{\smallskip} \hline \noalign{\smallskip}
			Lognormal (0,1) & 10 & 0.01 & 0.1883 & 0.0129 & 0.0053 & 0.0433 & 0.0026 & 0.0321 & 0.0847 \\
			& 20 &  & 0.6498 & 0.0445 & 0.0288 & 0.1878 & 0.0077 & 0.2372 & 0.5212 \\
			& 30 &  & 0.9129 & 0.0918 & 0.0763 & 0.4440 & 0.0309 & 0.6381 & 0.8835 \\
			& 40 &  & 0.9913 & 0.1577 & 0.1544 & 0.7057 & 0.2322 & 0.9599 & 0.9669 \\
			& 50 &  & 0.9990 & 0.2341 & 0.2666 & 0.8917 & 0.7072 & 0.9932 & 0.9888 \\
			\noalign{\smallskip} \hline \noalign{\smallskip}
			& 10 & 0.05 & 0.5036 & 0.0738 & 0.0560 & 0.1613 & 0.0247 & 0.1434 & 0.1977 \\
			& 20 &  & 0.9038 & 0.1856 & 0.1791 & 0.5317 & 0.0880 & 0.6309 & 0.6876 \\
			& 30 &  & 0.9934 & 0.2995 & 0.3368 & 0.7809 & 0.5245 & 0.9696 & 0.8050 \\
			& 40 &  & 0.9991 & 0.4397 & 0.5270 & 0.9321 & 0.8886 & 0.9953 & 0.8377 \\
			& 50 &  & 0.9999 & 0.5818 & 0.6795 & 0.9852 & 0.9893 & 0.9990 & 0.9004 \\
			\noalign{\smallskip} \hline \noalign{\smallskip}
			Pareto (1,1) & 10 & 0.01 & 0.0320 & 0.5922 & 0.7472 & 0.0179 & 0.1930 & 0.0580 & 0.6875 \\
			& 20 &  & 0.3753 & 0.9657 & 0.9925 & 0.0906 & 0.3853 & 0.2791 & 0.9002 \\
			& 30 &  & 0.9560 & 1.0000 & 1.0000 & 0.2576 & 0.5060 & 0.9481 & 0.9825 \\
			& 40 &  & 0.9997 & 1.0000 & 1.0000 & 0.6506 & 0.5863 & 0.9975 & 0.9967 \\
			& 50 &  & 1.0000 & 1.0000 & 1.0000 & 0.8538 & 0.6507 & 0.9998 & 0.9994 \\
			\noalign{\smallskip} \hline \noalign{\smallskip}
			& 10 & 0.05 & 0.2346 & 0.8486 & 0.9301 & 0.0911 & 0.2529 & 0.1959 & 0.2507 \\
			& 20 &  & 0.9244 & 0.9996 & 0.9997 & 0.3457 & 0.4417 & 0.8601 & 0.4055 \\
			& 30 &  & 0.9990 & 1.0000 & 1.0000 & 0.7700 & 0.5521 & 0.9982 & 0.5474 \\
			& 40 &  & 1.0000 & 1.0000 & 1.0000 & 0.9683 & 0.6401 & 1.0000 & 0.6518 \\
			& 50 &  & 1.0000 & 1.0000 & 1.0000 & 0.9968 & 0.6962 & 1.0000 & 0.8088 \\
			\noalign{\smallskip} \hline \noalign{\smallskip}
			Half-normal (2) & 10 & 0.01 & 0.0721 & 0.0126 & 0.0051 & 0.0219 & 0.0107 & 0.0164 & 0.1178 \\
			& 20 &  & 0.2334 & 0.0526 & 0.0267 & 0.0744 & 0.0351 & 0.0357 & 0.4864 \\
			& 30 &  & 0.3644 & 0.1450 & 0.0870 & 0.1508 & 0.0682 & 0.0736 & 0.6378 \\
			& 40 &  & 0.5151 & 0.2888 & 0.2122 & 0.2793 & 0.1098 & 0.1440 & 0.7804 \\
			& 50 &  & 0.6634 & 0.4551 & 0.3929 & 0.3801 & 0.1697 & 0.2598 & 0.8314 \\
			\noalign{\smallskip} \hline \noalign{\smallskip}
			& 10 & 0.05 & 0.2516 & 0.0804 & 0.0597 & 0.0966 & 0.0475 & 0.0690 & 0.1010 \\
			& 20 &  & 0.4698 & 0.2464 & 0.2117 & 0.2617 & 0.1057 & 0.1535 & 0.3646 \\
			& 30 &  & 0.6467 & 0.4642 & 0.4492 & 0.4342 & 0.1792 & 0.2365 & 0.5111 \\
			& 40 &  & 0.7885 & 0.6691 & 0.6873 & 0.5766 & 0.2670 & 0.4078 & 0.5997 \\
			& 50 &  & 0.8601 & 0.8151 & 0.8505 & 0.7171 & 0.4160 & 0.5595 & 0.6764 \\
			\noalign{\smallskip} \hline \noalign{\smallskip}
		\end{tabular} 
	\end{table}

					\subsection{Censored case}
					
					We calculate empirical type I error and power of the test statistic proposed for right censored data using Monte Carlo simulation studies. To calculate the empirical type I error, lifetimes are generated from Rayleigh distribution. We consider different choices of alternatives as in uncensored case for finding the empirical power. Here, the censoring  percentages  are chosen to be 20\% and 40\%.  In all cases, the censoring random variable $C$ is generated from exponential distribution with parameter $b$,  where $b$ is chosen such that $P(T>C)=0.2(0.4)$.  Re-weighting techniques explained in Section 3 is used to estimate the variance of $\widehat\Delta_c$. Results of the simulation study are presented in Table 3.
					
					
					\begin{table}[h]
						\caption{Empirical type I error and power of the test for censored data}
						\scalebox{0.8}{
							\begin{tabular}{cccccccccccccc}\hline
								\multirow{2}{*}{} & \multicolumn{2}{c}{Rayleigh (1)}& \multicolumn{2}{c}{Exponential (1)} & \multicolumn{2}{c}{Lognormal  (2,1)}&\multicolumn{2}{c}{Gamma (3,1)}&\multicolumn{2}{c}{Inv Gaussian (2,1)} \\ \hline
								$n$ &   $\alpha=0.01$  & $\alpha=0.05$&   $\alpha=0.01$  & $\alpha=0.05$ & $\alpha=0.01$  & $\alpha=0.05$&   $\alpha=0.01$  & $\alpha=0.05$&$\alpha=0.01$ &$\alpha=0.05$  \\ \hline
								\multirow{2}{*}{} & \multicolumn{10}{c}{20\% censoring}\\
								\hline
								50 &0.0149&0.0474&0.9891& 0.9967&0.8296& 0.8791&0.4794&0.8735& 0.8425& 0.9037\\
								75&0.0128&0.0488&  0.9925& 0.9974&0.6696& 0.8098 & 0.7907&0.9738& 0.9481& 0.9745\\
								100&0.0115&0.0491&1.0000& 1.0000& 0.9388& 0.9635& 0.9836&0.9998& 0.9917& 0.9986   \\
								200& 0.0107&0.0496& 1.0000& 1.0000& 0.9999&1.0000& 1.0000&  1.0000& 1.0000 &1.0000\\ \hline
								\multirow{2}{*}{} & \multicolumn{10}{c}{40\% censoring}\\
								\hline
								50 &0.0148&0.0524&0.7529&0.8403& 0.4366&0.5358& 0.1916& 0.6607&0.4978& 0.5812\\
								75 &0.0131& 0.0517& 0.9740& 0.9849&0.5444& 0.6476 & 0.3408& 0.8228 & 0.5759 & 0.6654  \\
								100&0.0118&0.0512&0.9895&0.9963& 0.6405&0.7346& 0.8028&0.9729&0.7758 &0.8326    \\
								200&0.0126&0.0507&0.9996& 1.0000&  0.6398&0.7618&0.9987&1.0000&70.9283&0.8326  \\ \hline
						\end{tabular}}
					\end{table}
					
					We can see that the  empirical power of the test approaches  the chosen level significance  as $n$ increases.  From Table 3, we observe that the power of the test  increases with sample size and decreases with censoring percentage.
					\section{Data analysis}
					The proposed test procedures are illustrated using several real data sets.
					\subsection{Uncensored data}
					We consider two data sets for the analysis. To find the critical region, we use  the following  algorithm.
					\begin{enumerate}
						\item Estimate the parameter of Rayleigh distribution from the observed data.
						\item Generate a random sample from Rayleigh distribution  using the estimated parameter in Step 1.
						\item Repeat Step 2, 10000 times and determine the simulated critical points. 
					\end{enumerate}
					{\bf Illustration 1}: To investigate the performance of the proposed test, we analyze the data, given in Caroni (2002), represents the failure times of 25 ball bearing . These failure times are: 17.88, 28.92, 33.00, 41.52, 42.12, 45.60, 48.48, 51.84, 51.96, 54.12, 55.56, 67.80, 67.80, 67.80, 68.64, 68.64, 68.88, 84.12, 93.12, 98.64, 105.12, 105.84, 127.92, 128.04, 173.40. Using the proposed methods, the test statistic is obtained as 0.4843 where the $1\%$ level critical values are  0.0863 and 0.8589, and the $5\%$ level critical values are  0.1916 and 0.7769 respectively. Hence accept the null hypothesis $H_0$, that the data on  failure times of ball bearings follow Rayleigh distribution. \\
					\noindent {\bf Illustration 2}: We also examine the time to breakdown of a electrical insulating fluid which was subject to a constant voltage stress of 32 kV. The data is studied in Lawless (2011) (page 3) and given in Example 1.1.5. We obtain the test statistic as -0.5643 where the $1\%$ level critical values are -0.0119 and 0.9396 and $5\%$ level critical value are 0.1109 and  0.8531. Hence we reject the null hypothesis that the data follows Rayleigh distribution.
					\subsection{Censored data}
					We consider two real data sets for illustration.  We use the normal based critical region given in Section 3 to make a decision.   The asymptotic null variance of $\widehat\Delta_c$ is estimated using the re-weighting techniques explained in Section 3. \\
					\noindent {\bf Illustration 1:} We examine the data on lifetimes of disk break pads on 40 cars studied in Lawless (2011). Complete data set is given in Table 6.11, Page 337 of Lawless (2011). Out of the 40 observed lifetimes, 9 are censoring times, hence data contains 22.5\% of censored observations. The test statistic is obtained as 0.7230. Hence we reject the hypothesis of Rayleigh distribution assumption for this data at both 1\% and 5\% level of significance.\\
					\noindent {\bf Illustration 2:} We analyse stanford heart transplant data available in R software  named `stanford2' to test for gamma assumption. The data consist of 184 lifetimes where 72 of them are censored lifetimes. The censoring percentage of data is 38.5\%. The test statistic is calculated as 50.2974. So we reject the null hypothesis that the data follows Rayleigh distribution at both 1\% and 5\% level of significance.
					
					\section{JEL ratio test}
					Empirical likelihood is a non-parametric inference tool based on likelihood principle  which originally developed by Thomas and Grunkemeier (1975) for obtaining the confidence interval for survival probability in the presents of right censored observations.  Pioneering papers by Owen (1988, 1990) for finding the confidence interval of regression coefficients and several statistical functionals  take the concept of empirical likelihood inference into a general methodology in statistical inference.  The implementation of  empirical likelihood inference is computationally challenging when the optimization of the non-parametric likelihood  has nonlinear constraints such as U-statistics with higher degree $(\ge 2)$ kernel.  To overcome this computational difficulties,  Jing et al. (2009) introduced the jackknife empirical likelihood (JEL) inference, which combines two popular non-parametric approaches, namely, the jackknife and the empirical likelihood method. Jing et al. (2009) illustrated the proposed methodology using one and two sample U-statistics. 
					
					Next we develop  jackknife empirical likelihood ratio  test for testing standard Rayleigh distribution.  For standard Rayleigh distribution the departure given in (\ref{delta}) become
					\begin{eqnarray}\label{deltams}
						\Delta_s(F)&=&\int_{0}^{\infty}\left(E\left[\left(\frac{x^2-1}{x}\right) min(X,t)\right]-F(t)\right)dF(t).
					\end{eqnarray}Hence (\ref{deltams}) become
					\begin{eqnarray}\label{deltafinal}
						\Delta_s(F)=E\left(\min(X_{1},X_{2})^2+X_1X_2I(X_2<X_1)\right)-E\left(\frac{X_2}{X_1}I(X_2<X_1)\right)-1.
					\end{eqnarray}Hence the U-statistics based  test is given by
					\begin{equation}\label{tests}
						\widehat{\Delta}_s=\frac{1}{n(n-1)}\sum_{i=1}^{n-1}\sum_{j=i+1}^{n}\left(2\min(X_{i},X_{j})^2+X_iX_j+\frac{X_{i}}{X_{j}}I(X_{i}<X_{j})
						+\frac{X_{j}}{X_{i}}I(X_{j}<X_{i})\right).
					\end{equation}
					In term of order statistics we can write above test statistic as
					\begin{equation}\label{tests1}
						\widehat{\Delta}_s=\frac{2}{n(n-1)}\sum_{i=1}^{n}(n-i)X_{(i)}^2+
						\frac{1}{n(n-1)}\sum_{j=1}^{n-1}\sum_{i=j+1}^{n}X_{(j)}\left({X_{(i)}}-\frac{1}{X_{(i)}}\right)-1.
					\end{equation}
					For developing JEL based goodness of fit test for Rayleigh distribution, first we construct jackknife pseudo values using the  test statistic given (\ref{tests}). Suppose $\widehat\Delta_{ks}$, are the value of the test statistic given in  (\ref{tests})  calculated    using $(n-1)$ observations  $X_1$, $X_2$,..., $X_{k-1}$,$ X_{k+1}$,...,$X_n$; $k=1,2,...,n$. The jackknife pseudo values denoted by $\nu_k$ are defined as
					$$\nu_k=n\widehat\Delta-(n-1)\widehat\Delta_{k},\,\, k=1,\ldots,n.$$
					Note  that $\nu_{k}$  are asymptotically independent under some mild conditions (Shi, 1984) and this  justify the use of $\nu_{k}$ in the empirical likelihood inference.
					
					Let $p=(p_1,\ldots,p_n)$ be a probability vector. It is well-known that $\prod_{i=1}^{n}p_i$ subject to $\sum_{i=1}^{n}p_i=1$ attain its maximum value $n^{-n}$ at $p_i=1/n$.
					Hence the jackknife empirical likelihood ratio for testing PED  based on the departure measure $\Delta_s(F)$ is  defined as
					$$R(\Delta)=\max\Big\{\prod_{i=1}^{n} np_i,\,\,\sum_{i=1}^{n}p_i=1,\,\,\sum_{i=1}^{n}p_i\nu_i=0\Big\},$$
					where
					$$p_i=\frac{1}{n}\frac{1}{1+\lambda \nu_i}$$
					and  $\lambda$ satisfies $$\frac{1}{n}\sum_{i=1}^{n}\frac{\nu_i}{1+\lambda \nu_i}=0.$$
					Hence the jackknife empirical log-likelihood ratio is given by $$\log R(\Delta)=-\sum\log(1+\lambda \nu_i).$$
					We find the limiting distribution of the jackknife empirical log-likelihood ratio to obtain the critical region of the JEL based test.  Using Theorem 1 of Jing et al. (2009) we have the following result as an analogue of Wilk's theorem.
					\begin{theorem}Let $h(X_1,X_2)=\left(\min(X_{1},X_{2})^2+X_1X_2+\frac{X_{1}}{X_{2}}I(X_{1}<X_{2})
						+\frac{X_{2}}{X_{1}}I(X_{2}<X_{1})\right)$. Suppose  $E(h^2(X_1,X_2))<\infty$ and $\sigma^2>0,$ then as $n\rightarrow\infty$,  $-2\log R(\Delta)$ converges in distribution to a $\chi^2$ random variable with one degree of freedom.
					\end{theorem}
					Using Theorem 5 we can obtain the critical region of the JEL based test. We reject the null hypothesis $H_0$ against the alternatives hypothesis $H_1$ at a  significance level $\alpha$, if
					\begin{equation}
						-2 \log R(\Delta)> \chi^2_{1,\alpha},
					\end{equation}	
					where $\chi^2_{1,\alpha}$ is the upper $\alpha$-percentile point of the $\chi^2$ distribution with one degree of freedom.
					\vspace{-0.5cm}
					\section{Concluding Remarks}
					Based on fixed point characterization for Rayleigh distribution, we developed new goodness of fit test for  Rayleigh distribution. We studied the asymptotic properties of the proposed test statistic. Monte Carlo simulation study is conducted to evaluate the finite sample performance of the test as well as to compare with other test available in literature.    The proposed test is illustrated using two real data sets. We also developed JEL ratio test for testing standard Rayleigh distribution.
					
					Even though many tests are available for Rayleigh distribution, as of our knowledge,  all of these test are developed for complete data. Motivated by this we develop a new  goodness of fit test for Rayleigh distribution under right censored data. We prove that the asymptotic distribution of the proposed test statistic is normal. We also find a consistent estimator of the asymptotic variance.  The finite sample performance of the test is evaluated through Monte Carlo simulation study.  Apart from right censoring, truncation and other types of censoring are common in lifetime data analysis. The proposed test can be modified to incorporates these situations. Goodness of fit tests for other important lifetime distributions can also be derived using similar fixed point characterizations. 
					\section*{Acknowledgements}
					Vaisakh K.M. and Sreedevi E.P. would like to thank Kerala State Council for Science, Technology and Environment for the financial support to carry out this research work. Thomas Xavier would like to thank Dr. Sudheesh K. and Dr. Isha Dewan for their support.


				\end{document}